\documentclass[prd,showpacs,preprintnumbers]{revtex4}
\usepackage{bm}

\begin{document}
\title{Quantum Electrodynamics based on a Superselection Rule} 
\author{Walter Smilga}
\affiliation{Isardamm 135 d, D-82538 Geretsried, Germany}
\email{wsmilga@compuserve.com}
\date{\today}

\begin{abstract}
This paper analyzes, for a multi-particle system of spin-1/2 particles, the 
consequences of replacing the Poincar\'e group as fundamental symmetry group 
by the de~Sitter group SO(3,2).
The flat-space approximation of the de~Sitter group by the Poincar\'e group  
defines a superselection rule, which correlates spin and momentum of 
particles. This correlation can be formulated as an interaction between two 
particles, which exhibits properties of the electromagnetic interaction.
\end{abstract}

\pacs{12.20.-m, 12.60.-i, 11.30.-j}

\maketitle

\section{Introduction} 

Assume that the basic symmetry group of particle physics is not the
Poincar\'e group P(3,1) but the de~Sitter group SO(3,2).
Let $l_{ab},\;a, b = 0,\ldots,4,$ be the generators of SO(3,2) 
pseudo-rotations within a representation of SO(3,2). 
Let $s^{ab}$ be 4x4-matrices, built 
from Dirac matrices, 
\begin{equation}
s_{\mu\nu} :=\, \frac{1}{2} \sigma_{\mu\nu}, 
\;\;
s_{4\mu} :=\, \frac{1}{2} \gamma_\mu\; ,\;\;\;\mu, \nu = 0,\ldots,3 
\label{0-6}
\end{equation} 
with the commutation and anticommutation relations
\begin{equation}
\sigma_{\mu\nu} = \frac{i}{2} [\gamma_\mu, \gamma_\nu]    	
\;\;\mbox{  and  }\;\;
\{\gamma_\mu, \gamma_\nu \} = 2 g_{\mu\nu}  \;.          \label{0-8}
\end{equation} 
$s_{ab}$ and $l_{ab}$ satisfy the commutation relations of SO(3,2) 
\begin{equation}
[l_{\mu\nu}, l_{\rho\sigma}] = 
-i[g_{\mu\rho}\, l_{\nu\sigma} - g_{\mu\sigma}\, 
l_{\nu\rho} + g_{\nu\sigma}\, l_{\mu\rho} 
- g_{\nu\rho}\, l_{\mu\sigma}] \mbox{ , } \label{0-2}
\end{equation}
\begin{equation}
[l_{4\mu}, l_{4\nu}] = -i g_{44} \; l_{\mu\nu} \;,       \label{0-3}
\end{equation}
\begin{equation}
[l_{\mu\nu}, l_{4\rho}] 
= i[g_{\nu\rho}\, l_{4\mu} - g_{\mu\rho}\, l_{4\nu}]   \label{0-4}	 
\end{equation} 
with $g_{ab}=$ diag $(+1,-1,-1,-1,+1)$.

Then the equation 
\begin{equation}
(2\,s^{ab} l_{ab} - m)\,\psi = 0  \label{0-1}
\end{equation}
(factor 2 is added for convenience)
is a SO(3,2) generalization of the Dirac equation.
The operators
\begin{equation}
j_{ab} = l_{ab} + s_{ab}                                 \label{0-9}
\end{equation}
correspond to the total pseudo-angular momentum of solutions of (\ref{0-1}).

The contraction limit \cite{iw} approximates SO(3,2) by P(3,1) in the
neighborhood of a given point in space-time.
The operators $l_{4\mu}$ are then approximated by momentum operators $p_\mu$
of P(3,1).
The operators $l_{\mu\nu}$ generate transformations of the common Lorentz 
subgroup SO(3,1) of SO(3,2) and P(3,1).

Consider now a multi-particle state space, formed by direct products of 
solutions of (\ref{0-1}). 
The total pseudo-angular momentum of the system is given by
\begin{equation}
J_{ab} = \sum{j_{ab}} \;.                                  \label{0-10}
\end{equation}
If the total state is a pure state, then the constant operator 
(Casimir operator) 
\begin{equation}
J^{ab} J_{ab} = \mbox{c-number}                            \label{0-11}
\end{equation}
defines a superselection rule. In the contraction limit it is replaced
by an analogous relation for the square of the total momentum $P_\mu$  
\begin{equation}
P^\mu P_\mu = M^2 \;.                                      \label{0-12}
\end{equation}

It will be shown that the contraction limit delivers a second superselection 
rule that correlates spin and momentum. It has its origin in invariant terms 
with the structure of $s^{ab}\,l_{ab}$, which is, not least, responsible 
for the existence of the generalized Dirac equation (\ref{0-1}).

\section{Group contraction}

Group contraction has been defined by E.~In\"on\"u and E.~P.~Wigner 
\cite{iw}, as a flat-space approximation for the neighborhood of a given 
point in space-time.
Consider the representation of the generators $l_{ab}$ by differential
operators acting on wave functions on the pseudo-sphere 
$g^{ab}\,x_a x_b = 1$
\begin{equation}
l_{ab} = i\, (x_a\, \frac{\partial}{\partial{x^b}} - x_b\, 
\frac{\partial}{\partial{x^a}})\;. \label{1-2}
\end{equation}
If, in a neighborhood of the point $P = (0, 0, 0, 0, 1)$, we rescale the 
coordinates $x_a$ by replacing
\begin{equation}
x_\mu = \frac{1}{R} \, \xi_\mu  \mbox{ and }  x_4 = \xi_4\;,   \label{1-3}
\end{equation}
(\ref{1-2}) can be written in the form
\begin{equation}
l_{4\mu} = i \, (R \, \xi_4 \, \frac{\partial}{\partial{\xi^\mu}} 
   - \frac{1}{R} \, \xi_\mu \, \frac{\partial}{\partial{\xi^4}}) \label{1-4} 
\end{equation}
and
\begin{equation}
l_{\mu\nu} = i \, (\xi_\mu \, \frac{\partial}{\partial{\xi^\nu}}
 - \xi_\nu \, \frac{\partial}{\partial{\xi^\mu}})\;.             \label{1-5} 
\end{equation}
For $R \to \infty$, the first term in (\ref{1-4}) is of order $R^1$, the
second of order $R^{-1}$ and $l_{\mu\nu}$ of order $R^0$.
When $R \to \infty$, any neighborhood of $P$, expressed in coordinates 
$\xi_\mu$, is `contracted' towards the point $P$.
At the same time the operator $l_{4\mu}$ is approximated by commuting 
operators
\begin{equation} 
p_\mu = i\,R\,\frac{\partial}{\partial{\xi^\mu}}\;.              \label{1-8}
\end{equation}
To get rid of the factor $R$, $p_\mu$ is rescaled while performing the 
limit.
This is done by replacing $\xi_\mu$ by $R\,\xi'_\mu$. 
According to (\ref{1-3}) the new coordinates $\xi'_\mu$ will be called 
$x_\mu$ again. But now $x_\mu$ are coordinates in tangential space-time. 
Then $p_\mu$ adopt the structure of generators of translations in 
tangential space-time
\begin{equation} 
p_\mu = i\,\frac{\partial}{\partial{x^\mu}}\;.                   \label{1-9}
\end{equation}

\section{Superselection rules}

With (\ref{1-8}) as an approximation to $l_{4\mu}$, we can arrange the terms 
of the Casimir operator (\ref{0-11}) with respect to their order of $R$. 
When we ignore terms that are of orders smaller than $R^1$, we obtain 
\begin{eqnarray}
& & p^\mu p_\mu + 2 p^\mu p'_\mu +  p'^\mu p'_\mu + \cdots \nonumber \\
&+& \gamma^\mu p_\mu + \gamma^\mu p'_\mu + \gamma'^\mu p_\mu + 
\gamma'^\mu p'_\mu + \cdots \nonumber \\ 
&=& \mbox{c-number} \;. 				\label{2-1}
\end{eqnarray}

In the first line, we find the Casimir operator $P^\mu P_\mu$ of the 
Poincar\'e group. 
In the second line, we find the contraction of the invariant form
$S^{ab}\,L_{ab}$, where $S^{ab} = \sum s^{ab}$ and $L_{ab} = \sum l_{ab}$.
It is evident that the first and the second line are separately 
invariant with respect to transformations of the Poincar\'e group.

Now consider a multi-particle system of spin-1/2 particles, whose individual
momenta add up to a total momentum $P$.
For this system $P^\mu P_\mu$ is a c-number. 
We can combine this number with the c-number on the right-hand side of 
(\ref{2-1}).  
Then we obtain another constant expression  
\begin{equation}
\gamma^\mu  (p_\mu  + p'_\mu + \cdots)   	
+ \gamma'^\mu (p'_\mu + p_\mu  + \cdots) 
+ \cdots
= \mbox{c-number} \;,		\label{2-4}	
\end{equation}
which is valid for all $R$, especially for $R \to \infty$,
up to contributions of orders less than $R^1$.  

In this way, the contraction limit produces two superselection rules. 
Rule I concerns, as expected, the constancy of the Casimir operator 
$P^\mu P_\mu$ of the Poincar\'e group. 
Rule II correlates spin and momentum within a multi-particle system.

For a single particle system, rule II reduces to the Dirac equation 
\begin{equation}
(\gamma^\mu p_\mu - m)\,\psi = 0 \;.    \label{2-5}
\end{equation}
Under the aspect of the basic SO(3,2) symmetry, (\ref{2-4}) is the correct 
extension of the single-particle Dirac equation to a multi-particle system.

\section{Interaction term in Fock space }

To further analyze the restrictions imposed by the correlation terms in 
(\ref{2-4}), we make use of standard Fock space methods, and treat the 
correlation terms as perturbation to the `free' parts of (\ref{2-4}). 
The `free' parts are defined by the terms $\gamma^\mu p_\mu$.
They are easily converted into a Fock space formulation, following the 
usual `second quantization' of the Dirac field. 
We refer to standard textbooks (see e.g. \cite{gs}).
The field operator of the Dirac field (taken from this reference) has
the form
\begin{equation}
\psi(x)\!=\!(2\pi)^{-\frac{3}{2}}\!\!\!\int\!d^3p 
\left( b_s({\mathbf{p}})u_s ({\mathbf{p}})e^{-ipx} 
\!+ {d^\dagger_s({\mathbf{p}})}v_s({\mathbf{p}})e^{ipx} \right).                                        \label{4-1}
\end{equation}
A similar expression defines the Dirac adjoint operator $\bar{\psi}(x)$. 
$b^\dagger_s({\mathbf{p}}), b_s({\mathbf{p}})$ are electron emission 
and absorption operators, $d^\dagger_s({\mathbf{p}}), d_s({\mathbf{p}})$ 
are the corresponding operators for positrons.
They satisfy the anticommutation relations of the Dirac field.

Unlike the usual approach to a multi-particle system of Dirac
particles, we are confronted with mixed terms that correlate the states 
of two particles at a time.
These correlation terms $\gamma^\mu p'_\mu$ of (\ref{2-4}) can be 
written (on the time-cut $t = t'$) as a Fock space operator 
\begin{equation}
\int\! d^3x\,d^3x' \; \bar{\psi}(x) \gamma^\mu \psi(x) \; 
\bar{\psi}(x') p'_\mu \psi(x') \;.                          \label{4-2}
\end{equation}
Using the decomposition of the field operators (\ref{4-1}) into emission
and absoption operators, the contributions to (\ref{4-2}) take on the form
\begin{equation}
\dots {b^\dagger({\mathbf{p+k}}) \, \gamma^\mu \, } {b({\mathbf{p}})\,\,} 
{b^\dagger({\mathbf{p'-k'}}) \, p'_\mu} \, b(\mathbf{p'})\dots \;. \label{4-3}
\end{equation}
(For simplicity we have omitted the factors $u_s$, $v_s$ and the time 
dependencies.)
When we evaluate (\ref{4-3}) for any two-particle state with a given total
momentum $P$, then only terms with $\mathbf{k=k'}$ will deliver a 
contribution.

If we use the correlation operator in a perturbation calculation,
then the requirement of momentum conservation is satisfied by only using
terms with $\mathbf{k=k'}$.
We can easily convince ourselves that any other term would violate 
momentum conservation.
Hence, the condition $\mathbf{k=k'}$ is required and sufficient to ensure 
momentum conservation in a perturbation calculation. 

Therefore, we are not only allowed to, but, in fact, are forced to drop all 
other terms in the decomposition (\ref{4-3}).
By collecting all contributions that belong to the same $\mathbf{p}$ 
and $\mathbf{k}$, we obtain 
\begin{equation}
\dots \; {b^\dagger({\mathbf{p + k}}) \, \gamma^\mu \, b({\mathbf{p}})} \, 
\kappa_\mu({\mathbf{k}}) \dots                        \label{4-4}
\end{equation}
with 
\begin{equation}
\kappa_\mu({\mathbf{k}}) := \int dV(p')
\,b^\dagger({\mathbf{p' - k}}) \, p'_\mu \, b({\mathbf{p'}})\;. \label{4-5}
\end{equation}
(Again time dependencies and spin functions are omitted.)
$dV(p')$ indicates a summation over all terms that contribute to a given 
$\mathbf{k}$.
The same consideration is valid for the positron operators 
$d^\dagger({\mathbf{p}})$, $d({\mathbf{p}})$ and mixed terms.

(\ref{4-4}) and (\ref{4-5}) are not yet suited for a standard perturbation
calculation. The reason is that this perturbation term is `nonlinear', in the
sense that it is defined as a product of two Fock space operators, built from 
the same field operators. 
In classical and quantum mechanics it is common practice to `linearize' a 
two-body problem by the introduction of a potential, which stands for the 
`forces' between the bodies.
The two-body problem is then reduced to the generally simpler one of a 
single body moving within a potential. 

(\ref{4-4}) and (\ref{4-5}) already have a form that strongly suggests, how 
a similar linearization can be achieved within our two-particle problem.
Consider the action of the operator $\kappa_\mu({\mathbf{k}})$ 
in (\ref{4-5}). It `absorbs' a particle with momentum $\mathbf{p'}$ and 
`recreates' this particle with momentum $\mathbf{p' - k}$.
In this way a momentum $\mathbf{k}$ is `emitted'.
$\kappa_\mu({\mathbf{k}})$ then acts as a placeholder for $\mathbf{k}$
and `transfers' this momentum to (\ref{4-4}), where it is 
`absorbed'\footnote{Of course, no physical creation or annihilation of 
particles is associated with these processes.}.

In the following we will reproduce these features by a quantum mechanical 
potential.
The essential step will be, to split up the aforementioned 
emission/absorption process, and assign specific emission and absorption
operators $a^\dagger_\mu({\mathbf{k}})$ and $a_\mu({\mathbf{k}})$
to each part of the process.
These operators, applied to the same vacuum state as 
$b({\mathbf{p}})$, $b^\dagger({\mathbf{p}})$,
shall emit and absorb quanta of momentum $\mathbf{k}$. 
As such they have to satisfy the commutation relations 
\begin{equation}
[a_\mu({\mathbf{k}}), a^\dagger_\nu ({\mathbf{k'}})] 
= \delta_{\mu\nu}\,\delta({\mathbf{k - k'}}) \;.             \label{4-7}
\end{equation}
Just as $\kappa_\mu({\mathbf{k}})$ they shall transform like a 
4-momentum.  

These operators will take over the placeholder role in (\ref{4-4}),
if we make sure that, together, they act in the same way as 
$\kappa_\mu({\mathbf{k}})$.
This is achieved by the requirement that, under inclusion of the 
placeholder momentum $\mathbf{k}$, energy-momentum is conserved at each 
`vertex' (\ref{4-4}). 
The commutation relations (\ref{4-7}) finally ensure that any combination
of $a_\mu({\mathbf{k}})$ and $a^\dagger_\mu({\mathbf{k'}})$ that, on the 
time-cut, does not satisfy the requirement $\mathbf{k=k'}$, is eliminated 
by multiplication to the vacuum state, either on the right or left side
of a matrix element.

By this step we have linearized the perturbation term by adding a
new `vector field' generated by $a_\mu({\mathbf{k}})$, 
$a^\dagger_\mu({\mathbf{k}})$.

Below we will also use the following operators, known from the conventional 
formulation of quantum electrodynamics (see e.g. \cite{gs}), 
\begin{equation}
A_j(x) = (2\pi)^{-3/2} \!\!\int\!\frac{d^3k}{\sqrt{2 k^0}} 
\left( a_j({\mathbf{k}})\,e^{-ikx} 
+ a^\dagger_j({\mathbf{k}})\,e^{ikx} \right)\;, 
\label{4-8}
\end{equation}
$j=1,2,3,$ and
\begin{equation}
A_0(x) = (2\pi)^{-3/2} \!\!\int\!\frac{d^3k}{\sqrt{2 k^0}} 
\;i \left( a_0({\mathbf{k}})\,e^{-ikx} 
+ a^\dagger_0({\mathbf{k}})\,e^{ikx} \right)\;.  \label{4-9}
\end{equation}
$k^0$ shall be determined by the requirement of energy-momentum conservation, 
when these operators are evaluated within two-particle states. 
This leads to an `off-shell' behavior, which is well-known from standard
perturbation theory\footnote{In the `free radiation field' $k^0$ is 
`on-shell': $k^0 = |\mathbf{k}|$.}.  

In (\ref{4-4}) we now add the proper space-time dependencies to the emission 
and absorption operators. 
By replacing $\kappa_\mu$ in (\ref{4-4}) by $e\,a_\mu$ we write
\begin{equation}
\dots \; e \; b^\dagger({\mathbf{p+k}})\, e^{i(p+k)x}\; \gamma^\mu \;
b({\mathbf{p}})\, e^{-ipx}\,\, 
a_\mu({\mathbf{k}})\, e^{-ikx} \dots \;.                 \label{4-10}
\end{equation}
$e$ is a normalization factor that must be added, since (\ref{4-7})
fixes a normalization of 
$a_\mu({\mathbf{k}})$, $a^\dagger_\mu({\mathbf{k}})$,
which cannot be expected to be the same as of $\kappa_\mu({\mathbf{k}})$.
Notice that the correct space-time dependency of $a_\mu({\mathbf{k}})$ is 
determined by (\ref{4-5}), when the proper time dependencies are added to 
$b^\dagger({\mathbf{p' - k}})$ and $b ({\mathbf{p'}})$.

After inserting the spin functions $u_s(\mathbf{p})$ and $v_s(\mathbf{p})$,
these terms and the corresponding positron and mixed terms, in 
combination with $a_\mu({\mathbf{k}})$ and $a^\dagger_\mu({\mathbf{k}})$,
add up to a Fock space operator in the form
\begin{equation}
e \int d^3x\, : \bar{\psi}(x)\gamma^\mu \psi(x) : A_\mu(x)\;. \label{4-11}
\end{equation}
Here $::$ stand for normal ordering of emission and absorption operators
(all emission operators left of all absorption operators).
This operator has the form of the interaction term of quantum 
electrodynamics (QED) with a coupling constant~e.

\section{ Discussion}

We have found that, within the framework of perturbation theory, the 
interaction term of QED is identical to a Fock space representation of the 
correlation terms of (\ref{2-4}).
Therefore, quantum electrodynamics can be considered as a formulation in
Fock space of superselection rule II.
Since QED is central to particle physics, 
this strongly suggests that the Hilbert space of particle physics has a
basic symmetry defined by the de~Sitter group SO(3,2), rather than by the 
Poincar\'e group\footnote{The symmetry of space-time is a different story.}. 
This symmetry can be approximated with high precision, but not replaced, by a 
Poincar\'e symmetry.
As a consequence of the SO(3,2) symmetry, spin-1/2 particles exhibit an 
inherent property of electromagnetic interaction.

In deriving the interaction term, we have linearized a two-body problem
by introducing a quantum mechanical potential.
The potential has not been obtained by a formal `quantization rule', 
applied to a classical potential, but rather by explicit construction on the 
quantum mechanical level. 
This provides valuable insights into the mechanism of `interacting quantized 
fields'. 

The perturbation algorithm of QED calculates transition amplitudes between 
`incoming' states ($t \rightarrow -\infty$) and `outgoing' 
states ($t \rightarrow \infty$).
Our derivation of the interaction term leads to the following interpretation
of its function within this algorithm: 
Applied to incoming states with momenta $\mathbf{p}$ and $\mathbf{p'}$, it 
develops an \begin{em}entangled\end{em} two-particle state. 
This state comprises all combinations of individual momenta that add up 
to the same value of $\mathbf{p+p'}$.
The entanglement is built up in such a way that correlations between spin 
and momentum of \begin{em}different\end{em} particles are established, 
in compliance with super\-selection rule II. 
By a measurement, which means a projection onto `outgoing' states, 
transition amplitudes are obtained. These connect the incoming states to 
other combinations within the entangled two-particle state.

Notice that entanglement basically is a non-local phenomenon. 
Nevertheless, after the introduction of the potential, the interaction term 
is strictly local.
This is a direct consequence of energy-momentum conservation at the `vertex'.

In our approach `intermediate photons' appear as `pseudo-particles',  
manifesting correlations within a multi-particle system.
They are auxiliary elements within an algorithm, introduced ad hoc to simplify
the algorithm. 
These elements describe correlation, but defi\-nitely not the physical 
`creation' and `annihilation' of photons as independent physical 
entities\footnote{This is not in contradiction to the experimental 
`observation' of photons: we do observe the interaction of a photon with a 
detector, but never a photon `itself'.}.

Consequently, there is also no `virtual pair production' from `intermediate
photons'.
Actually, there are the same commutation functions as in the conventional 
formulation of QED, so-called `contractions', which in Feynman graphs are 
symbolized by photon and fermion lines. 
They are responsible for the correct transfer of energy-momentum through 
the iterated interaction term, with the important consequence of relativistic 
causality in the space-time domain.
Within our approach they are obtained in a transparent way as part of the 
perturbation algorithm, which does not leave room for any other interpretation.
This means, there is no `metamorphosis' of plain Fock space operators
to `physical fields', when the interaction term `is switched on'. But the most 
significant consequence is this: After the interaction `is switched on', 
a two-fermion system is still a two-fermion system and can be completely 
described and understood within the framework of elementary Fock space 
methods\footnote{Many textbooks tell us the contrary, namely that interacting 
relativistic field theories always deal with an unlimited number of particles.
The reason for this opinion is a misinterpretation of Fock space operators, 
taking the words `emission' and `absorption' too literal.
In the past we have too readily adopted this interpretation, obviously because
it mirrors what we believe to experience in `daily live'.}.

The consequences for the physical interpretation of some typical Feynman 
diagrams were discussed in \cite{ws}.
In the same reference, an estimate of the numerical value of the 
coupling constant $e$ was obtained by evaluating the volume element 
$dV(p')$ of (\ref{4-5}). 
It reproduces Wyler's formula \cite{aw} of the fine-structure constant.

Ignoring contributions to (\ref{0-11}) of orders smaller than $R^1$ 
means ignoring terms with the structure of $\gamma^\mu \gamma'_\mu\,$ and 
$\,x^\mu \partial^4 \, \partial'_\mu\,$.
An tentative interpretation of these highly interesting terms was given 
in~\cite{ws}.

\end{document}